\documentstyle[pre,preprint,aps,epsfig,amsmath]{revtex}

\newcommand{\beq}{\begin{equation}}
\newcommand{\eeq}{\end{equation}}
\newcommand{\bea}{\begin{eqnarray}}
\newcommand{\eea}{\end{eqnarray}}
\newcommand{\barr}{\begin{array}}
\newcommand{\earr}{\end{array}}

\draft 
\begin{document}


\title{ {\bf \large Driven granular gases with gravity}}

\author{        
  A. Baldassarri$^{1}$,
  U. Marini Bettolo Marconi$^{1}$,
  { A. Puglisi$^2$}, 
  and A. Vulpiani$^2$}

\address{(1) Dipartimento di Matematica e Fisica, Universit\`a di Camerino,
Via Madonna delle Carceri, I-62032 , Camerino, Italy and
Istituto Nazionale di Fisica della Materia, Unit\`a di Camerino}

\address{(2) Dipartimento di Fisica, Universit\`a La Sapienza,
Piazzale A. Moro 2, 00185 Roma, Italy and
 Istituto Nazionale di Fisica della Materia, Unit\`a di Roma}

\maketitle

\begin{abstract}


We study fluidized granular gases
in a stationary state determined by the balance between an external
driving and the bulk dissipation.
The two considered situations are inspired by recent
experiments, where the gravity plays a major role as a driving
mechanism: in the first case gravity acts only in one direction and
the bottom wall is vibrated, in the second case gravity acts in both
directions and no vibrating walls are present. 
Simulations performed under the molecular chaos assumption 
show averaged profiles of density, velocity and
granular temperature which are in good agreement with the
experiments. Moreover we measure the velocity distributions which
show strong non-Gaussian behavior, as experiments pointed out, but also
density correlations accounting for clustering, at odds with the
experimental results.  The hydrodynamics of the 
first model is discussed and an exact solution is found for the density
and granular temperature as functions of the distance from the
vibrating wall. The limitations of such a solution, in particular 
in a broad layer near the wall injecting energy, are discussed.
\end{abstract}

\pacs{81.05.Rm, 05.20.Dd}


\section{Introduction}

In general granular materials~\cite{general}, since the presence of
dissipative forces, are not equilibrium systems neither from the
configurational point of view nor from the dynamical point of view.  A
statistically stationary state can be produced by the competition
between the dissipation due to the inelastic collisions among the
particles and the energy injection due to an external source which
prevents the system from cooling and come to rest.
 
Usually granular gases are considered in the homogeneous cooling
regime, less frequently they are studied in a stationary regime where
energy flows into the system from some external source (stochastic
driving, vibrating plates, shear, etc.) and dissipates by means of
inelastic collisions. A sufficient condition to prevent strong density
instabilities (such as those found by Du et al.~\cite{kadanoff}),
seems to be the presence of an even minimal, but spread, temperature
source~\cite{nostri}.

Many evidences, by mean of computer simulations, have been found that
different kinds of density instabilities, like {\em
clustering}~\cite{zanetti} (density gradients growing on time scales
faster than typical hydrodynamics scales) or {\em inelastic
collapse}~\cite{mcnamara} (the local divergence of collision rate so
that an infinite number of collisions occurs in a finite time) may
emerge in a cooling granular assembly, that is a granular gas losing
his starting kinetic energy because of dissipative collisions. It has
also been shown that the velocity distribution of particles in the
free cooling state with homogeneous density has overpopulated high
energy tails $\sim exp(-Av)$~\cite{poschel,olandesi}.

When granular gases are driven in some way to balance the loss of
energy due to collisions, a stationary state may be observed. The
first model of randomly driven granular gas was proposed
in~\cite{kadanoff}. It showed pathologies in the density and granular
temperature profiles but also a breakdown of thermodynamic
limit. Another randomly driven model was then proposed to offer a
different insight into the kinetics of granular
gases~\cite{nostri}. In this model the driving mechanism is
a stochastic energy source acting on every particle as a heat bath with a fixed
temperature $T_F$ and a fixed viscous damping with characteristic time
$\tau$. In the stationary ``collisional'' regime (characterized by a
collision time much lower than $\tau$) the gas showed a fractal
distribution of density and a distribution of velocities with
overpopulated (non-Gaussian) high energy tails. The homogeneous
solution of the corresponding Boltzmann-Enskog equation has been
analytically studied~\cite{olandesi} showing that $\sim
exp(-Av^{3/2})$ high energy tails are expected. 

The aim of this work is to study a class of models for driven granular
gases where the efficiency of the energy injection is guaranteed by
the presence of gravity, taking inspiration by some recent
experiments~\cite{kudrolli,azanza}: in these experiments a bottom
confining wall is the source of granular temperature while gravity
forces the particles to return in contact with this source.  
We are interested in very diluted systems, where the granular
material behaves as an inelastic gas, rather than dense granular
flows, where many static effects, such as clogging, arching or
bubbling appear. Such systems have been studied in relation with
compaction dynamics or slow dense chute flows\cite{densi}.
The study
is based on Direct Simulation Monte Carlo, but we also discuss (for
one of the models) the hydrodynamic theory.  The first version of the
model (gravity in only one direction and vibrating bottom wall) has
been previously studied in the one-dimensional case, that is a
vibrated column of grains under the force of gravity~\cite{1d} and the
transition or the coexistence of different phases (gas, partially
fluidized and condensed) was investigated. In two dimensions
experiments~\cite{warr}, simulations~\cite{luding} and
theories~\cite{kumaran} have analyzed a vertical system of grains with
gravity and a vibrating bottom wall (with different kinds of
vibration) searching for a simple scaling relation between global
variables as the global granular temperature $T_G$ or the center of
mass height $h_{cm}$ as function of the size of the system $N$, the
typical velocity of the vibrating wall $V$ or the restitution
coefficient $r$. In all these calculations the authors did not pay too
much attention to the hydrodynamic profiles of the system, always
assuming a constant granular temperature (``isotherm atmosphere'') and
a density profile exponentially decaying with the height, as in the
case of a Boltzmann elastic gas under gravity. One of the results of
this work, discussed in section V, is that also in the dilute regime
that one can study by means of Monte Carlo methods the use of these
assumptions is not obvious, in particular when trying to solve the
global balance between external energy injection and bulk dissipation
due to inelastic collisions among particles. It must be also noted
that the general validity of a hydrodynamical description is still
subject of debate in the case of granular gases far from the elastic
limit (a review of hydrodynamical problems is~\cite{goldhirsch}).

In section II we present the two versions of the model, in section III
and IV we illustrate the results, in section V we discuss the
hydrodynamics  of the model in its first
 version, and finally we draw the conclusions.
For the sake of completeness and in order to make the paper self-contained
we included in Appendix A a brief description of the 
Direct Simulation Monte Carlo of the Boltzmann equation and in
Appendix B the expressions of the dimensionless coefficients appearing
in the hydrodynamic equations of section V.


\section{The Models}

We introduce two bi-dimensional 
models both consisting of $N$ identical
smooth disks of diameter $\sigma$ and mass $m=1$ subject to binary instantaneous inelastic
collisions  which conserve the total momentum
\beq
{\bf v}_1'+{\bf v}_2'={\bf v}_1+{\bf v}_2
\end{equation}
and reduce the normal component of the relative velocity 
\beq
({\bf v}_1'-{\bf v}_2') \cdot {\bf \hat{n}}=
-r(({\bf v}_1-{\bf v}_2) \cdot {\bf \hat{n}})
\end{equation}
where $r$ is the normal restitution coefficient ($r=1$ in the completely
elastic case) and ${\bf \hat{n}}=({\bf x_1}-{\bf x_2})/\sigma$ is the 
unit vector along the line of centers ${\bf x}_1$ and ${\bf x}_2$ of the colliding disks at
contact. With these rules satisfied the post-collisional
velocities are:

\bea
{\bf v_1}'={\bf v}_1-\frac{1+r}{2}(({\bf v}_1-{\bf v}_2) \cdot {\bf
\hat{n}}){\bf \hat{n}} \nonumber \\
{\bf v}_2'={\bf v}_2+\frac{1+r}{2}(({\bf v}_1-{\bf v}_2) \cdot {\bf
\hat{n}}){\bf \hat{n}} 
\end{eqnarray}

\begin{figure}[h]
	\centerline{\psfig{figure=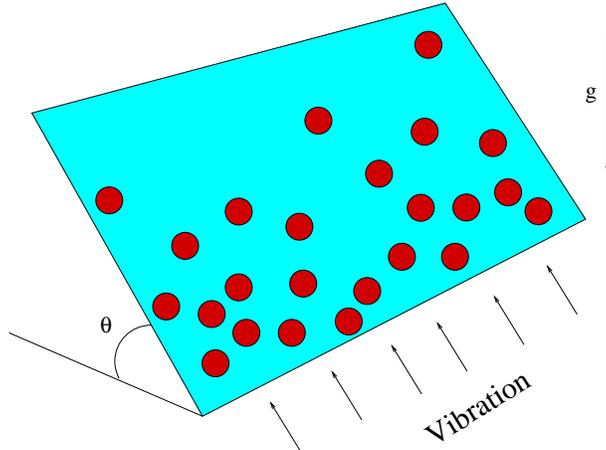,angle=-90,width=8cm}}
	\caption{A sketch of the first model where the granular
	assembly is driven by gravity plus a (periodically or
	stochastic) vibrating wall}	
	\label{fig:a_sketch}
	\end{figure}

\begin{figure}[h]
	\centerline{\psfig{figure=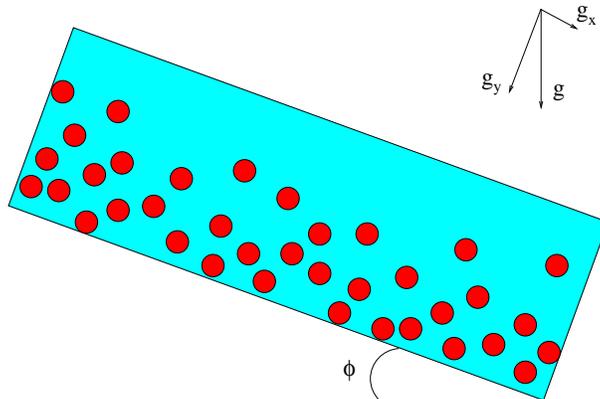,angle=-90,width=8cm}}
	\caption{A sketch of the second model where the only energy
	source is gravity, with components in both directions}
	\label{fig:b_sketch}
	\end{figure}

In addition, the particles experience the external gravitational field
and the presence of confining walls. With respect to previous works~\cite{nostri} 
the energy necessary to prevent the cooling of the system due to the 
inelastic collisions is not provided by a heat bath: in the present 
paper the energy feeding mechanism is of two types according to
the two numerical experiments we perform.
\begin{itemize}
\item
In model a), illustrated in fig. (\ref{fig:a_sketch}) and inspired to a
recent laboratory experiment~\cite{kudrolli} and a
numerical experiment~\cite{giappone}, the ``apparatus''
consists of a plane of dimension $L_x \times L_y$ inclined by an angle
$\theta$ with respect to the horizontal. The particles are
constrained to move in such a plane under the action of an 
effective gravitational force $g_e=g\sin{\theta}$ pointing
downward. 
In the horizontal direction there are periodic
boundary conditions. Vertically the particles are confined by walls,
both inelastic with a restitution coefficient $r_w$ (difference of
restitution coefficients for particle-particle interactions and
particle-wall interactions will be discussed below). Besides, the
bottom wall vibrates and therefore injects energy
and momentum into the system.  
The vibration can have either a periodic
character (as in~\cite{kudrolli}) or a stochastic behavior with thermal
properties (as in~\cite{giappone}). In the
periodic case, the wall oscillates vertically with the
law $Y_w=A_w \sin(\omega_w t)$ and the particles collide with it as with
a body of infinite mass, so that the vertical component of their
velocity after the collision is $v_y'=-r_wv_y+(1+r_w)V_w$ where
$V_w=A_w \omega_w \cos(\omega_w t)$ the velocity of the  vibrating wall. 
In the stochastic case we assume
that the vibration amplitude is negligible and that the particle
colliding with the wall have, after the collision, new random
velocity components $v_x \in (-\infty,+\infty)$ and $v_y\in
(0,+\infty)$ with the following probability
 distributions: 

\begin{equation} P(v_y)=\frac{v_y}{T_w}exp(-\frac{v_y^2}{2 T_w}) \end{equation}

\begin {equation} P(v_x)=\frac{1}{\sqrt{2 \pi T_w}}exp(-\frac{v_x^2}{2 T_w}) \end{equation}

\item
In model b) sketched in fig. (\ref{fig:b_sketch}) the ``set-up'' 
is a two dimensional channel of depth $L_y$ and of length $L_x$, vertically confined by a
bottom and a top inelastic wall,  with periodic boundary
conditions in the direction parallel to the flow. The channel is tilted up
by an angle $\phi$ with respect to the horizontal so that gravity
has both components $g_x=g\sin{\phi}$ and $g_y=g\cos{\phi}$.
This model mimics the experiment performed by Azanza {\em et
al.}~\cite{azanza}, where a stationary flow in a two dimensional inclined
channel was observed at a point far from the source of the granular material. 
The assumption of periodic boundary conditions in the direction of the
flow is consistent with the observed stationarity, due to the
balance between the gravity drift and the damping effect of inelastic
collisions (for a discussion of the possible regimes that can be
shown by one particle in presence of this balance, see~\cite{umberto}).
\end{itemize}

The chosen collision rule excludes the presence of tangential forces,
and hence the rotational degrees of freedom do not contribute to the
description of the dynamics.

Under the assumption of {\em molecular chaos},
stating that 
$P_2({\bf x},{\bf x}+\sigma{\bf \hat{n}},{\bf v}_1,{\bf v}_2,t)=P({\bf x},{\bf v}_1,t)
P({\bf x}+\sigma{\bf \hat{n}},{\bf v}_2,t)$ 
where $P_2$ and $P$ are the probability density functions 
for two particles and one particle respectively, it is possible to
write down the Boltzmann equation (eq. (\ref{boltzeq}) in
section V), which can be solved by means of Monte Carlo methods.
Here we used a simplified (but
still efficient) version of the Direct Simulation Monte Carlo scheme
proposed by Bird~\cite{bird}. With respect to the
original version of the algorithm the
clock which determines the collision rate is replaced by an
a-priori fixed collision rate via a constant collision probability $p_c$ given
to every disk at every time-step $\Delta t$ of the simulation, in such
a way that the single-particle collision rate is $\chi \sim
p_c/\Delta t$. The colliding particle then seeks its
collision partner among the other particles in a neighborhood of radius
$r_B$, choosing it randomly with a probability proportional to their
relative velocities.  Moreover in this
approximation the diameter $\sigma$ is no more explicitly relevant but
it is directly related to the choices of $p_c$ and $r_B$ in a non
trivial way: in fact the Bird algorithm allows the particles to pass
through each others, so that a precise diameter cannot be defined and estimated
as a function of $p_c$ and $r_B$. The Bird scheme is described in more
details in the Appendix A.

The agreement between our simulations and the inspiring experiments,
justifies the simplifying assumptions considered for our model, i.e.
assuming molecular chaos and neglecting tangential
forces.  Nevertheless, as a partial check, we try a modified version of
model b) where the tangential forces may affect the post-collisional
velocities of the particles. As reported below, the introduction of
such forces does not change the behavior of the measured
quantities.


\section{Discussion of the results: \\ model a)}

Simulations of the first model, the inclined plane with a bottom wall
injecting energy, have been performed for different choices of 
the number of disks $N$, the normal restitution coefficient $r$,
the dimensionless width of the plane $N_w=L_x/r_B$  and the parameter
measuring the rate of energy injection from the wall, that is the temperature
$T_w$ in the stochastic case and the amplitude and frequency $A_w$,
$\omega_w$ in the periodic case.  

Let us show how  numerical simulations with
the molecular chaos assumption reproduce the main results obtained in
experiments~\cite{kudrolli,azanza} and in high performance
computer simulations~\cite{giappone} of inelastic hard disks.

Snapshots of the systems and time-averaged density profiles are shown
 in fig. (\ref{fig:tflip_dens}) for the case of randomly vibrated wall. 
We are in the presence of a highly fluidized phase of the type Isobe
 and Nakanishi~\cite{giappone}
 call granular
 turbulent: looking at the time evolution of the
density distribution of the system and of the coarse-grained velocity
field one observes an intermittency-like behavior with rapid and strong
fluctuations of the density, sudden explosions followed by large clusters
of particles traveling downward, coherently, under the action of
 gravity. Of course more dense and ordered phases (that one can expect
 at lower values of energy injection) are not
 reproducible with the Direct Simulation Monte Carlo, as strong
 excluded volume effects appear and the
 assumption of negligible short range correlations fails.

\begin{figure}[h]
	\centerline{\psfig{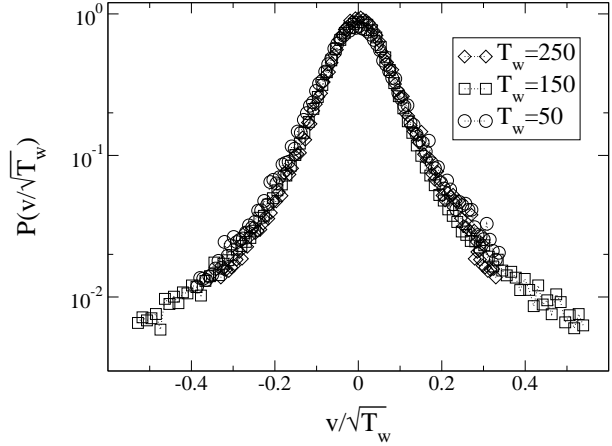}}
	\caption{Distribution of rescaled horizontal velocities
	$v/\sqrt{T_w}$ for the
	model a) with stochastic wall at different temperatures
	$T_w=50$, $T_w=100$, $T_w=250$. The other parameters are 
	$N=5000$, $N_w \approx 180$, $r=0.7$, $r_w=0.7$, $g_e=-1$}	
	\label{fig:tflip_vglob}
	\end{figure}

\begin{figure}[h]
	\centerline{\psfig{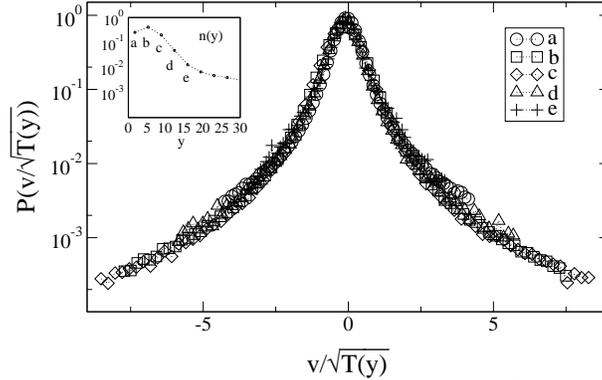}}
	\caption{Distribution of horizontal velocities, for the model 
	a) with stochastic wall, measured on stripes at different
	heights and rescaled by the average temperature at that
	height. The inset shows the normalized number density profile with the position
	of the chosen stripes. $N=5000$, $N_w
	\approx 180$, $r=0.7$, $r_w=0.7$, $g_e=-1$, $T_w=100$}	
	\label{fig:tflip_vstripe}
	\end{figure}

\begin{figure}[h]
	\centerline{\psfig{figure=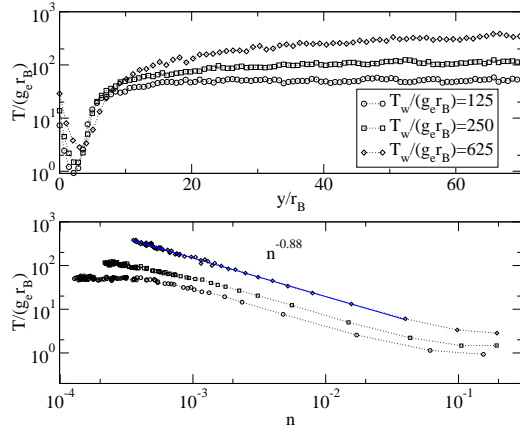,angle=-90,width=8cm}}
	\caption{Granular (dimensionless) temperature $T/(g_e r_B)$
	versus dimensionless height
	$y/r_B$ (above) and versus number density $n$ (bottom) for the for the model 
	a) with stochastic wall, with $N=5000$, $N_w \approx 180$,
	$r=0.7$, $r_w=0.7$, $g_e=-1$. The solid line is a power-law
 	fit for $T(n)$.}	
	\label{fig:tflip_t}
	\end{figure}

In the figures (\ref{fig:tflip_vglob}) and (\ref{fig:tflip_vstripe}) 
we display the horizontal velocity
distributions, for the stochastic case. In fig. (\ref{fig:tflip_vglob})
 distributions for different $T_w$ are shown: the data collapse is obtained
by rescaling the velocities by
$\sqrt{T_w}$. 
Instead, in fig. (\ref{fig:tflip_vstripe}) we show the velocity distributions of
particles contained in stripes at different heights from the wall,
again rescaled by $\sqrt{T(y)}$ (their own variance) in order to
obtain the data collapse. It
appears that the distributions are non-Gaussian and their broadening
(that is the granular temperature $T(y)$) is density dependent. This
dependence is shown in fig. (\ref{fig:tflip_t}) as well as its dependence upon the height.
An analogous dependence has been shown in references~\cite{nostri}
where the granular gas was driven by a homogeneous heat bath, showing
a power law $T \sim n^{-\beta}$ with $\beta \sim 0.8$, while in this
case it seems $\beta \sim 0.88$.

\begin{figure}[h]
	\centerline{\psfig{figure=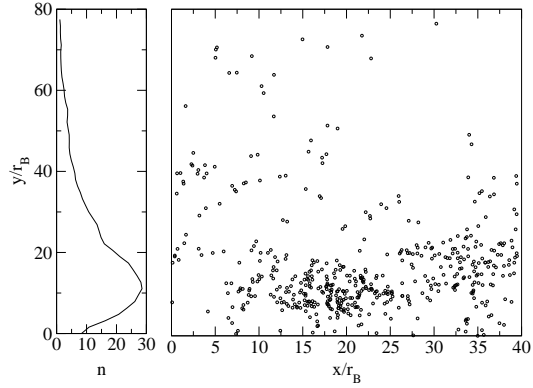,angle=-90,width=8cm}}
	\caption{Snapshot of the model a) with periodically vibrating
	wall (right) and time-averaged density profile (left) for the
	following choice of parameters: $N=500$, $N_w \approx 56$,
	$r=0.5$, $r_w=0.7$, $g_e=-1$, $f_w=400\pi$, $A_w=0.1$}
	\label{fig:vflip_dens}
	\end{figure}

\begin{figure}[h]
	\centerline{\psfig{figure=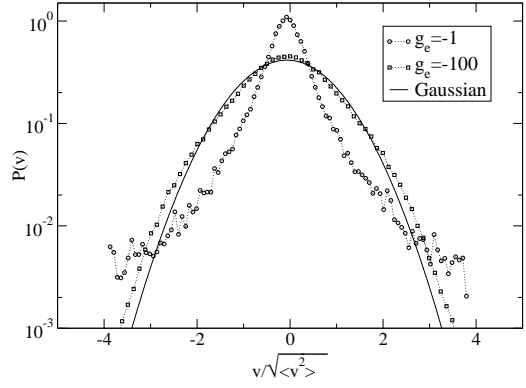,angle=-90,width=8cm}}
	\caption{Distributions of horizontal velocities for the model
	a) with periodically vibrating wall for two different values
	of inclination, that is $g_e=-1$ and $g_e=-100$, while
	the other parameters are fixed: $N=500$, $N_w \approx 56$,
	$r=0.5$, $r_w=0.7$, $f_w=400 \pi$, $A_w=0.1$}
	\label{fig:vflip_v}
	\end{figure}

The case of periodically vibrated wall is illustrated in figures
 (\ref{fig:vflip_dens}) and (\ref{fig:vflip_v}).
One can see the density profiles (together with a snapshot of
the system) and the distribution of horizontal velocities in two
different regimes: for $g_e=-1$ a non-Gaussian distribution is
obtained, while a distribution close to a Gaussian appears when
$g_e=-100$. This trend towards a Gaussian, as
the angle of inclination is raised up, reproduces exactly the
experimental observation of Kudrolli and Henry~\cite{kudrolli} (where
the angle of inclination of the plane was raised up from $\theta=0.1^o$
to $\theta=10^o$) and can
be explained as an effect of the increase of the collision rate with the
wall which ``randomizes'' the velocities in a more efficient way: this
resembles the heath bath model~\cite{nostri} where one passes from the
non-Gaussian regime to the Gaussian one increasing the ratio between the
heating rate and the collision rate. 

\begin{figure}[h]
	\centerline{\psfig{figure=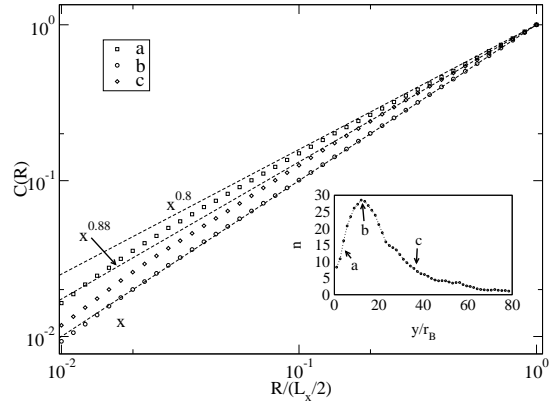,angle=-90,width=8cm}}
	\caption{Cumulated correlation function C(R), as defined in	
	the text, measured along stripes at different heights for the
	model a), with periodically vibrating wall. In the
	inset is displayed the number density profile, with the
	position of the chosen stripes. Here $N=500$, $N_w \approx 56$,
	$r=0.5$, $r_w=0.7$, $f_w=400\pi$, $A_w=0.1$ and $g_e=-1$}
	\label{fig:vflip_sc}
	\end{figure}

In order to characterize the spatial
clustering we have studied the cumulated particle-particle
correlation function:

\beq 
C_{B(y,\Delta y)} (t,R)=\frac{1}{N_{B(y,\Delta y)}(N_{B(y,\Delta y)}-1)}\sum_{i \neq
j: {\bf x}_i,{\bf x}_j \in B(y, \Delta y)}\Theta(R-|{\bf x}_i(t)-{\bf
x}_j(t)|)
\end{equation} 

where $B(y, \Delta y)$ is a horizontal stripe contained between $y+\Delta
y/2$ and $y-\Delta y/2$. After having checked that the system has
reached a stationary regime, we have computed the time-average of the
correlation function, that is 

\beq
C_{B(y,\Delta y)} (R)=
\frac{1}{T-t_0}\int_{t_0}^T dt C_{B(y,\Delta y)} (t,R)
\end{equation}

which is independent on time if $T>>t_0$. In the figure (\ref{fig:vflip_sc}) we show
the $C(R)$ vs. $R$  for different stripes $B(y, \Delta y)$.
We observe a power law behavior

\begin{equation}
C_{B(y,\Delta y)}(R) \sim R^{d_2(y)}
\end{equation}

In the case of homogeneous density $d_2$ is expected to be the
topological dimension of the stripe, that is $d_2=1$ if $R\gg\Delta y$
and $d_2=2$ if $R\ll\Delta y$.

Clustering, whose signature is a value of the 
correlation dimension $d_2$ lower than the topological dimension,
appears in the stripes with not too high densities, where an exponent
smaller than $1$ is measured (the fit is performed in th region $R\gg
\Delta y$). The evidence of
clustering is at odds with the observation of Kudrolli and
Henry~\cite{kudrolli}: 
they report, in fact, the absence of clustering by measuring 
the distribution of the number of particles in boxes of fixed
dimensions spread all over the inclined plane. This observation is
perhaps due to the fact that in the statistical analysis employed
in ref. \cite{kudrolli} the number of particles in each box is
considered disregarding their heights, that is they may belong 
to regions of different
densities: in such a way the slow decaying tails,  expected for
the clusterized distributions of the stripes at lower densities, are
partially hidden by the Poissonian (homogeneous) distribution of 
the stripes at
higher density. Moreover, even from the global density distribution
measured in their work, a tail decaying slower than a Poissonian
cannot be clearly ruled out.


\section{Discussion of the results: \\  model b)}

Let us now show the results for the second model, the inclined
bidimensional channel.

\begin{figure}[h]
	\centerline{\psfig{figure=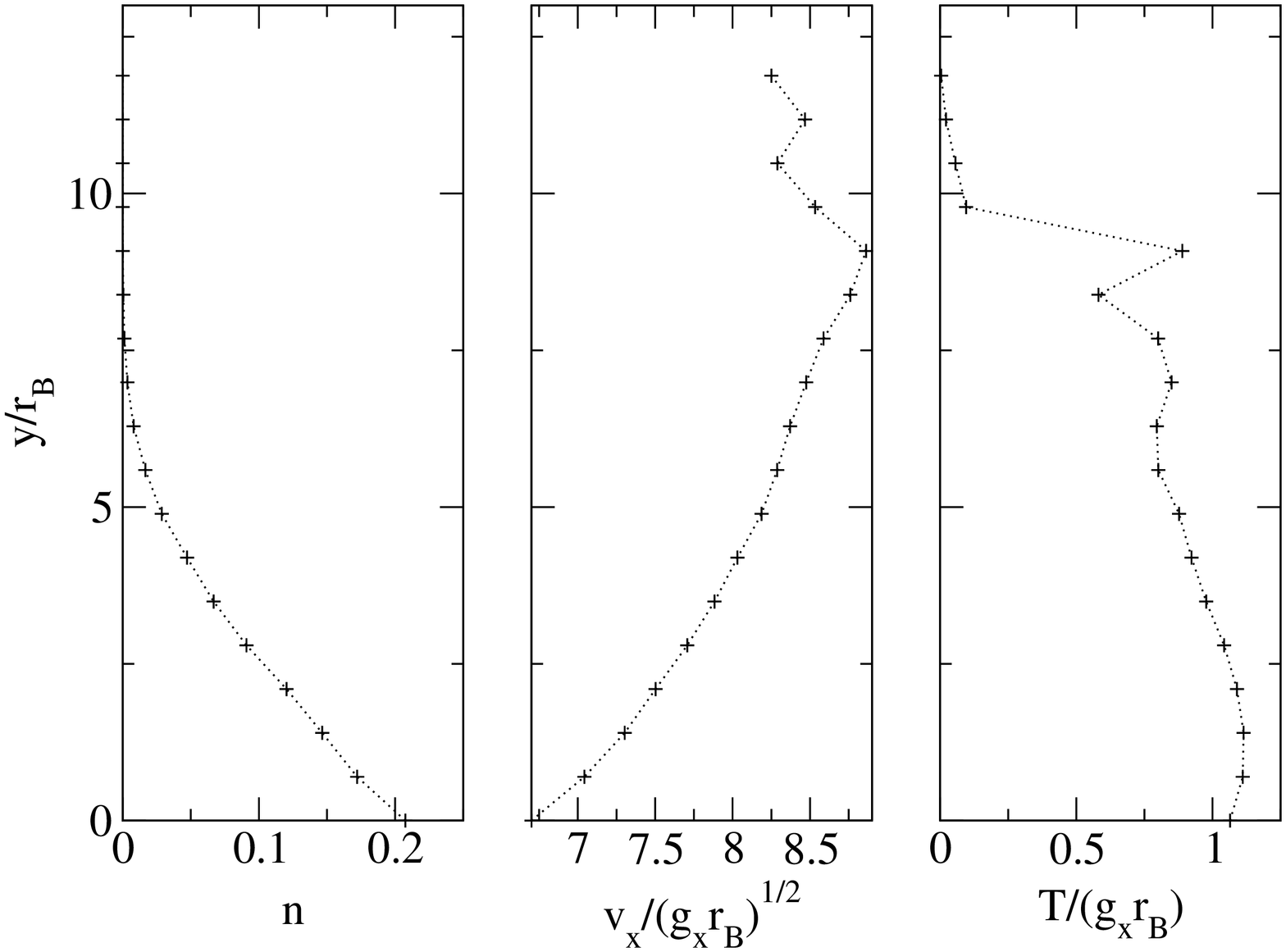,angle=-90,width=8cm}}
	\caption{Normalized number density $n$, dimensionless horizontal velocity
	$v_x/\sqrt{g_xr_B}$ and dimensionless granular
	temperature $T/\sqrt{g_xr_B}$ versus dimensionless height 
	$y/r_B$ for the two dimensional inclined channel (model b):
	$N=500$, $N_w \approx 56$, $g_x=1$, $g_y=-2$ (i.e.: the inclination angle
	$\phi=\pi/6$), $r=0.95$, $r_w=0.95$}
	\label{fig:c_prof1}
	\end{figure}

\begin{figure}[h]
	\centerline{\psfig{figure=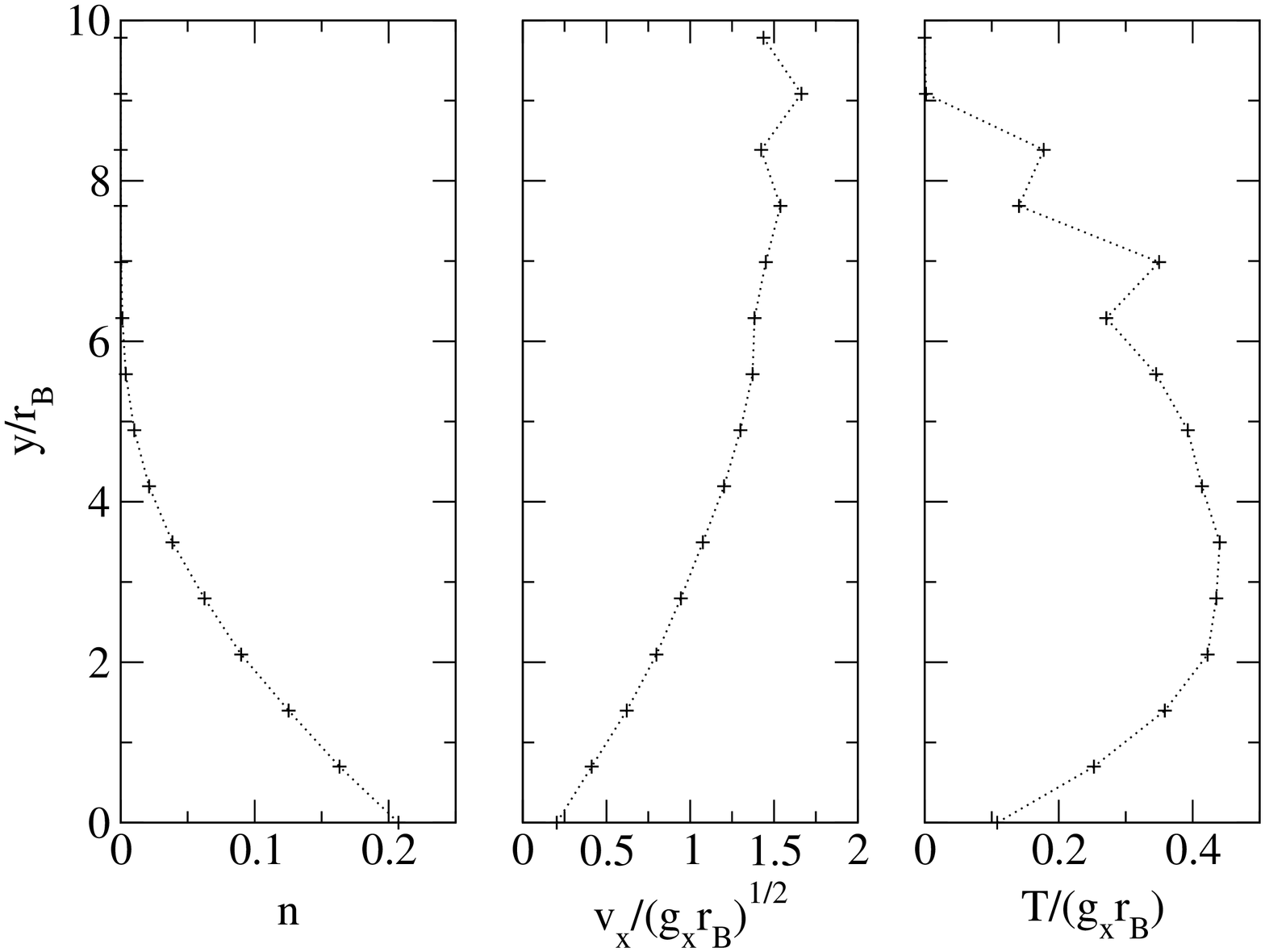,angle=-90,width=8cm}}
	\caption{Normalized number density $n$, dimensionless horizontal velocity
	$v_x/\sqrt{g_xr_B}$ and dimensionless granular
	temperature $T/\sqrt{g_xr_B}$ versus dimensionless height 
	$y/r_B$ for the two dimensional inclined channel (model b):
	$N=500$, $N_w \approx 56$, $g_x=1$, $g_y=-2$ (i.e.: the inclination angle
	$\phi=\pi/6$), $r=0.95$, $r_w=0.4$}
	\label{fig:c_prof2}
	\end{figure}

In figures (\ref{fig:c_prof1}) and
(\ref{fig:c_prof2}) the hydrodynamical fields $n(y)$ (number density),
$v_x(y)$ (velocity component parallel to the flow), $T(y)$ (granular
temperature) are shown as functions of the distance from the bottom
wall $y$. The velocity, the temperature and the height are made
dimensionless by rescaling them by $\sqrt{g_xr_B}$, $g_xr_B$ and $r_B$
respectively.  
The profiles well reproduce those measured experimentally by Azanza
{\em et al.}~\cite{azanza}: they show a critical height $H$ of about
six times the radius $r_B$ which corresponds to the separation between
two different regimes of the cooling rate. In a mean field framework
the local rate of dissipation due to the inelastic collisions (as
already stated before) is $\zeta \propto nT^{3/2}$.

\begin{figure}[h]
	\centerline{\psfig{figure=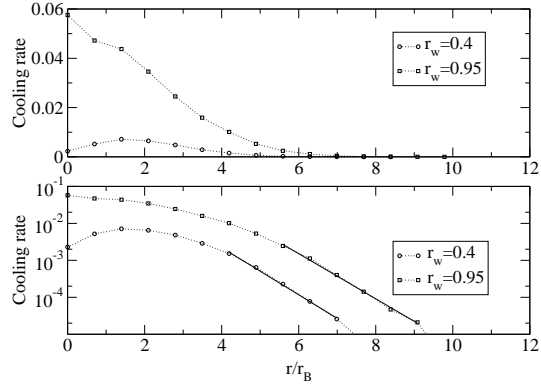,angle=-90,width=8cm}}
	\caption{Cooling rate, as defined in the text, versus
	dimensionless height $y/r_B$ for the two dimensional inclined channel (model b):
	$N=500$, $N_w \approx 56$, $g_x=1$, $g_y=-2$ (i.e.: the inclination angle
	$\phi=\pi/6$), $r=0.95$, $r_w=0.95$ or $r_w=0.4$}
	\label{fig:c_cool}
	\end{figure}

This can be understood simply noting that the collision rate is
proportional to the local density and to the local relative velocity
of the particles ($\sqrt{T}$), while the change in the granular
temperature induced by every collision is proportional to the
temperature $T$. The quantity $\tilde{\zeta}=nT^{3/2}$ as a function
of $y$ is shown in fig. (\ref{fig:c_cool}). The cooling rate decreases
exponentially and is reduced under $1/100$ of its maximum value at
about the observed critical height $H \approx 6r_B$, accounting for
the difference between a collisional regime and a ballistic one.


With respect to the velocity and temperature profiles in figure
(\ref{fig:c_prof1}), we note here that quite unphysical features
appear: in particular the quite strong slipping effect near the bottom
wall is in contrast with the experimental findings. We think that this
is due to a wrong modelling of the particle-wall collision events.

The restitution coefficient used in our model has to be considered as
an effective parameter describing the energetics of collisions.  It
should depend on the details of the collision event, in principle even
on the relative velocities of the colliding particles.  In the
experiment the bottom wall was covered with particles identical to the
flowing ones with a spacing bounded between $0$ and $0.8$ mm: however
the particles are stuck to the bottom wall so that the collision event
is completely different from a two-particles collision.

Using a lower effective restitution coefficient for the wall $r_w$
(see figure (\ref{fig:c_prof1})) we obtain a better agreement with the
experimental profiles.  In particular, both temperature and velocity
profiles seems to go to zero near the bottom, although we cannot
really rule out slipping effects ($v_x(y=0)\neq 0$).

\begin{figure}[h]
	\centerline{\psfig{figure=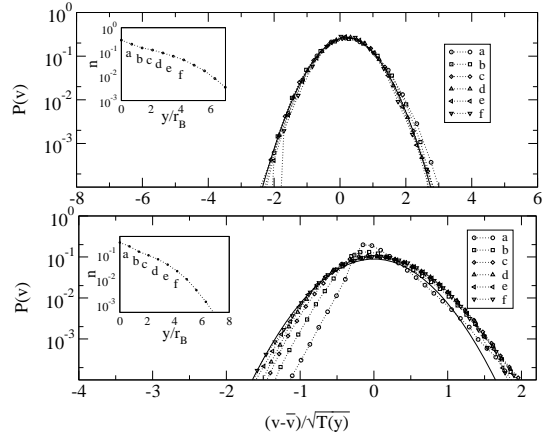,angle=-90,width=8cm}}
	\caption{Distribution of horizontal velocities for the model 
	b), measured on stripes at different
	heights and rescaled in order to have the same mean and
	variance. The inset shows the normalized number density profile with the position
	of the chosen stripes. $N=500$, $N_w \approx 56$,
	$r=0.95$, $r_w=0.95$, $g_x=1$, $g_y=-2$ (i.e.: the inclination angle
	$\phi=\pi/6$)}	
	\label{fig:c_v}
	\end{figure}

We have also studied the distribution of horizontal velocities in
stripes at different heights (here the mean values are height
dependent). These are displayed in fig. (\ref{fig:c_v}) showing the
emergence of a non-Gaussian behavior mainly in the case with $r_w<r$
and only in the stripes near the bottom wall. The
authors of the experiment of ref.\cite{azanza} claim that the
distributions of velocity are very close to the Gaussian and try to fit their data
with the rheological model proposed by Jenkins and
Richman~\cite{old_hydro},  which
postulate a quasi-Gaussian equilibrium 
to calculate the transport coefficients. Near the bottom wall
Gaussian approximation is far from obvious, as shown by the results of
our simulations: this is an effect of the inelasticity of the collisions
but also of the proximity of the boundary.

\begin{figure}[h]
	\centerline{\psfig{figure=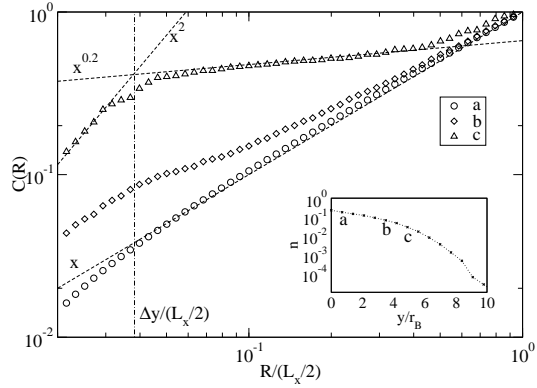,angle=-90,width=8cm}}
	\caption{Cumulated correlation function C(R), as defined in
	the text, measured along stripes at different heights for the
	model b). In the
	inset is displayed the normalized number density profile with the
	position of the choose stripes. Here $N=500$, $N_w \approx 56$,
	$r=0.95$, $r_w=0.95$, $g_x=1$, $g_y=-2$ (i.e.: the inclination angle
	$\phi=\pi/6$). The dashed lines represent the power-law
	fits, the vertical dot-dashed line represent the width of the
	stripes $\Delta y$}
	\label{fig:c_sc}
	\end{figure}

Finally we have investigated the homogeneity of the density: the
figure (\ref{fig:c_sc}) shows the previously defined function
$C_{B(y,\Delta y)}(R)$ for stripes at different density. It appears
again a clustering effect, with a correlation dimension ranging from
$1$ (homogeneous stripes) to $0.2$ (highly clusterized stripes).  In
the figure it is also shown the very small distance region, $R<r_B$,
where homogeneity should be recovered. Since in our simulation,
$\Delta y \approx r_B$, we expect $d(y)=2$ in this region.

We consider the comparison between our simplified model and the
experimental profiles quite satisfactory: this seems to suggest that
introducing further physical details should be irrelevant at this
description level. However we briefly report the results obtained with
a slightly modified version of the model, including the effects of
tangential forces. Such forces play a key role in dense granular
flows\cite{densi,forces}, being responsible for arching. On the other hand
the present results suggest that in the case of diluted systems they
act similarly to the normal forces without introducing noticeable
effects.
  
The introduction of tangential forces in the model studied accounts
for a new collision rule:
\begin{eqnarray}
({\bf v}_1'-{\bf v}_2') \cdot {\bf \hat{n}}&=&
-r^n(({\bf v}_1-{\bf v}_2) \cdot {\bf \hat{n}})\nonumber\\
({\bf v}_1'-{\bf v}_2') \cdot {\bf \hat{t}}&=&
-r^t(({\bf v}_1-{\bf v}_2) \cdot {\bf \hat{t}})\nonumber
\end{eqnarray}
where we replace the single restitution coefficient with a pair of
parameters $r^n$ and $r^t$, respectively due to the effect of normal
and tangential collision forces ($\hat{t}$ is a  unit
vector perpendicular to  $\hat{n}$). Analogously, the restitution
coefficient $r_w$ splits in two new parameters $r_w^n$ and $r_w^t$.
The results of simulations with several choices of the enlarged set of
parameters do not show qualitative differences: setting tangential
restitution coefficients lower than one is equivalent to enhance the
dissipation in the original model.

\begin{figure}[h]
        \centerline{\psfig{figure=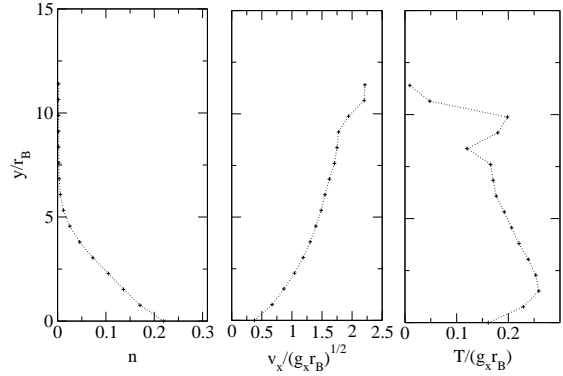,angle=-90,width=8cm}}
        \caption{Normalized number density $n$, dimensionless
        horizontal velocit y $v_x/\sqrt{g_xr_B}$ and dimensionless
        granular temperature $T/(g_xr_B)$ versus dimensionless height
        $y/r_B$ for the two dimensional inclined channel. Here
        tangential restitution coefficients smaller than one are
        considered (see text): $N=500$, $N_w \approx 56$, $g_x=1$,
        $g_y=-2$ (i.e.: the inclination angle $\phi=\pi/6$),
        $r^n=0.95$, $r^t=0$, $r_w^n=0.95$,$r_w^t=0$}
        \label{fig:c_prof1_bis} \end{figure}

Just as an example of this, we show figure (\ref{fig:c_prof1_bis}),
where  the extremal case of a vanishing tangential restitution
coefficient is reported. Note that the profiles are similar to those
shown in figure  (\ref{fig:c_prof2}) where a low $r_w=0.4$ was used.



\section{Discussion of the hydrodynamics: results and problems}

The Boltzmann equation for the two models introduced in this paper (in
two dimensions) reads:

\begin{equation}
\left( \frac{\partial}{\partial t} + {\bf v} \cdot \nabla+ g_i
\frac{\partial}{\partial v_i} \right) f({\bf x},{\bf v},t)=J(f,f)
\label{boltzeq} 
\end{equation}

\begin{equation}
J(f,f)=\sigma \int d{\bf v}_1 \int d{\bf \hat{n}} \Theta({\bf
\hat{n}} \cdot {\bf v}_r)({\bf
\hat{n}} \cdot {\bf v}_r)[r^{-2}f({\bf x},{\bf v}',t)f({\bf x},
{\bf v}_1',t)-f({\bf x},{\bf v},t) f({\bf x},{\bf v}_1,t)] 
\end{equation}

Here ${\bf \hat{n}}$ is the unit vector along the line joining
the centers of the colliding particles at contact, ${\bf v}_r={\bf
v}-{\bf v}_1$ is the relative velocity of the colliding disks, $\Theta$ is the Heaviside
step function, ${\bf v}'$ and ${\bf v}_1'$ are the precollisional
velocities leading after collision to velocities ${\bf v}$,${\bf
v}_1$.

The equation (\ref{boltzeq}) must be completed with the boundary
conditions in order to describe the microscopic evolution of the whole system.

The difficulty of solving the Boltzmann equation (\ref{boltzeq}) can
be bypassed substituting the microscopic description given by $f({\bf
x},{\bf v},t)$ with the averaged macroscopic description given by the
hydrodynamic fields: the number density field $n({\bf
x},t)$, the velocity field ${\bf v}({\bf x},t)$ and the granular
temperature field $T({\bf x},t)$. These quantities are given by

\begin{equation}
n({\bf x},t)=\int d{\bf v}f({\bf x},{\bf v},t)
\end{equation}

\begin{equation}
{\bf u}({\bf x},t)=\frac{1}{n({\bf x},t)}\int d{\bf v}{\bf v}f({\bf x},{\bf v},t)
\end{equation}

\begin{equation}
k_BT({\bf x},t)=\frac{1}{n({\bf x},t)}\int d{\bf v}\frac{m({\bf v}-{\bf u({\bf x},t)})^2}{2}f({\bf x},{\bf v},t)
\end{equation}

Multiplying the Boltzmann equation (\ref{boltzeq}) by $1$ or ${\bf v}$
or $m({\bf v}-{\bf u({\bf x},t)})^2/2$ and integrating over ${\bf v}_1$ one can
derive~\cite{chapman,cercignani} the equations of fluid dynamics:

\beq \frac{D n}{Dt}+ n \partial_i u_i=0 \label{h1}\end{equation}

\beq m n \frac{D u_i}{Dt}=-\partial_j \tau_{ij}+n g_i m\label{h2}\quad (i=1,2,3)
\end{equation}

\beq  n \frac{Dk_BT}{Dt}=- \partial_i q_i
-\tau_{ij}\partial_ju_i -\zeta nk_BT\label{h3}\end{equation}

where $\partial_i=\partial/\partial x_i$ (for the sake of compactness
 we use here the notation $x \rightarrow x_1$ and $y \rightarrow x_2$)
 and  $D/Dt=\partial/\partial
 t+{\bf u}\cdot {\bf \nabla}$ is the
Lagrangian derivative, e.g.: $\frac{D}{Dt} F({\bf x},t)=\frac{d}{dt}
F({\bf \phi}({\bf x}_0,t),t)$ with ${\bf \phi}({\bf x}_0,t)$ the
evolution after a time $t$ of ${\bf x}_0$ under the velocity field
${\bf u}$. In the above equations 

\begin{equation}
\tau_{ik}=\int d{\bf v} m (v_i-u_i)(v_k-u_k) f({\bf x},{\bf v},t)
\end{equation}

is the stress tensor, ${\bf g}$ is the volume external force (gravity
in our case),

\begin{equation}
q_i=\int d{\bf v} \frac{m}{2}v_i |{\bf v}-{\bf u}|^2 f({\bf x},{\bf v},t)
\end{equation}

is the heat flux vector and 

\begin{equation}
\zeta({\bf x},t)=\frac{m(1-r^2)\pi^{1/2} \sigma}{8 \Gamma(5/2)nk_BT} \int d{\bf
v}_1\int d{\bf v}_2 |{\bf v}_1-{\bf v}_2|^3f({\bf x},{\bf v}_1,t)f({\bf x},{\bf v}_2,t)
\end{equation}

is the cooling rate due to dissipative collisions.

The set of equations (\ref{h1})-(\ref{h3}) become closed hydrodynamic equations
for the fields $n$, ${\bf u}$ and $T$ when $P_{ij}$, ${\bf q}$ and $\zeta$ are
expressed as functionals of these fields. This is obtained, for
example, expressing the space and time dependence of $f$ in terms of
the hydrodynamic fields and then expanding $f$ to first order (the
so-called Navier-Stokes order) in their gradients, with the exception
of $\zeta$ which requires an expression of $f$ to the second order of
gradients to be consistent with the other terms. With this
approximation the equations (\ref{h1})-(\ref{h3}) include the contributions up
to the second order in the gradients of the fields. 

Calculations of the closure of the hydrodynamic equations for granular
media have been performed with some approximations restricting
the validity of the results to the low dissipation or quasi-elastic
limit~\cite{old_hydro}. More recently~\cite{brey} the analysis has
been extended to arbitrary inelasticity giving closed expressions for
the momentum and heat fluxes and for the cooling rate $\zeta$.

We follow these more recent results~\cite{brey} and write down the
hydrodynamics for the model a) presented in this paper (gravity in one
direction and vibrating bottom wall, i.e. ${\bf g}=(0,g_e)$ and
$g_e<0$ ), with the following assumptions: the fields do not depend
upon $x$ (the coordinate parallel to the bottom wall),
i.e. $\partial/\partial x=0$, and the system is in a steady state,
i.e. $\partial/\partial t=0$. The continuity equation (\ref{h1}) then
reads $\frac{\partial}{\partial y} (n(y)u_y(y))=0$ and this can be
compatible with the bottom and top walls (where $nv_y=0$) only if
$n(y)v_y(y)=0$, that is in the absence of macroscopic vertical flow.
The equations are written for the dimensionless fields
$\tilde{T}=k_BT/(-g_em\sigma)$ and $\tilde{n}=n\sigma^2$, while the
position $y$ is made dimensionless using $\tilde{y}=y/\sigma$. Finally
for the pressure we put $p(y)=\tau_{22}=n(y)k_BT(y)$. With the
assumption discussed above the equations of Brey et al.~\cite{brey}
read:

\begin{equation}
\frac{d}{d\tilde{y}}(\tilde{n}(\tilde{y})\tilde{T}(\tilde{y}))=-\tilde{n}(\tilde{y})
\label{hb1}
\end{equation}

\begin{equation}
\frac{1}{\tilde{n}(\tilde{y})}\frac{d}{d\tilde{y}}Q_r(\tilde{y})+C(r)\tilde{n}(\tilde{y})\tilde{T}(\tilde{y})^{3/2}=0
\label{hb2}
\end{equation}

where $Q_r(\tilde{y})$ is the granular heat flux expressed by

\begin{equation}
Q_r(\tilde{y})=A(r)\tilde{T}(\tilde{y})^{1/2}\frac{d}{d\tilde{y}}\tilde{T}(\tilde{y})+B(r)\frac{\tilde{T}(\tilde{y})^{3/2}}{\tilde{n}(\tilde{y})}\frac{d}{d\tilde{y}}\tilde{n}(\tilde{y})
\label{heatflux}
\end{equation}

In the above equations $A(r)$, $B(r)$ and $C(r)$ are dimensionless
monotone coefficients, all with the same sign, explicitly given in
the Appendix B. In particular $B(1)=0$ and $C(1)=0$, i.e. in the
elastic limit there is no dissipation and the heat transport is due
only to the temperature gradients, while when $r<1$ a term dependent
upon $\frac{d}{d\tilde{y}}\ln(\tilde{n}(\tilde{y}))$ appears in $Q_r(\tilde{y})$.
The use of dimensionless fields eliminates the explicit $\bf{g}$
dependence from the equations, that remains hidden in their structure
(the right hand term of equation \ref{hb1}, that is due to the gravitational
pressure gradient, disappears in the equation for $g=0$).

A change of coordinate can be applied to
eqs. (\ref{hb1}),(\ref{hb2}) in order to obtain a simpler form:

\begin{equation}
\tilde{y} \rightarrow l(\tilde{y})=\int_0^{\tilde{y}} \tilde{n}(y')dy'
\label{change}
\end{equation}

It follows that when $y$ spans the range $[0,L_y]$, the coordinate $l$
spans the range $[0,\sigma/L_x]$. With this change of coordinate it
happens that 

\begin{equation}
\frac{d}{d\tilde{y}} \rightarrow \tilde{n}(l)\frac{d}{dl}
\end{equation}

and the first equation (\ref{hb1}) reads:

\begin{equation}
\frac{d}{dl}(\tilde{n}(l)\tilde{T}(l))=-1
\label{hb1bis}
\end{equation}

from which is immediate to see that 

\begin{equation}
H=\tilde{n}(l)\tilde{T}(l)+l
\label{constant}
\end{equation}

is a constant, i.e. $\frac{d}{dl}H=0$. This is equivalent to observe
that 

\begin{equation}
P(y)-g\int_0^yn(y')dy'
\end{equation}

is constant which is nothing more than the Bernoulli theorem for a
fluid in the gravitational field with the density depending upon the height.

\begin{figure}[h]
	\centerline{\psfig{figure=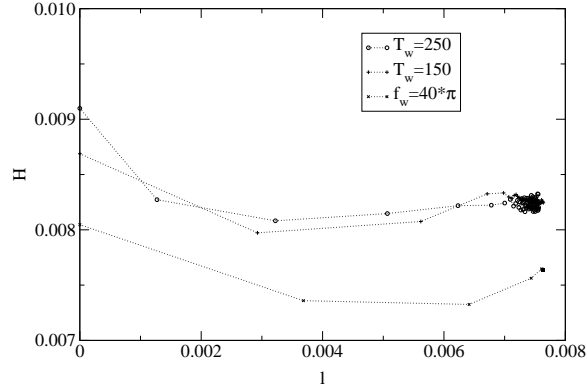,angle=-90,width=8cm}}
	\caption{Plot of $H$, defined in section V, versus $l$, for
	three different simulations of the model a): two cases are
	with the stochastic wall ($N=5000$, $N_w \approx
	180$, $r=0.7$, $r_w=0.7$, $g_e=-1$, $T_w=150$ and
	$T_w=250$), while the third case is with the periodic wall
	($N=5000$, $N_w \approx 180$, $r=0.7$, $r_w=0.7$, $g_e=-1$,
	$f_w=80\pi$, $A_w=0.1$}
	\label{fig:h}
	\end{figure}

The relation (\ref{constant}) is verified by the model
simulated in this work in the figure (\ref{fig:h}) for almost
all the height of the container, apart of the boundary layer near the
bottom driving wall.

Using the coordinate $l$ introduced in (\ref{change}) and the elimination
of $\tilde{n}(l)$ using the recognized constant, that is  

\begin{equation}
\tilde{n}(l)=\frac{H-l}{\tilde{T}(l)}
\label{substitution}
\end{equation}

the second equation (\ref{hb2}), after some simplifications, and after
a second change of coordinate $l \rightarrow s(l)=H-l$, becomes:

\begin{equation}
\frac{\alpha(r) s}{\tilde{T}(s)^{1/2}} \frac{d^2}{ds^2}\tilde{T}(s)-
\frac{\alpha(r)s}{2\tilde{T}(s)^{3/2}}\left(\frac{d}{ds}\tilde{T}(s)\right)^2-
\frac{\beta(r)}{\tilde{T}(s)^{1/2}}\frac{d}{ds}\tilde{T}(s)+s\tilde{T}(s)^{1/2}=0
\label{hb2bis}
\end{equation}

where $\alpha(r)=(A(r)-B(r))/C(r)=$, $\beta(r)=(A(r)-\frac{1}{2}B(r))/(C(r))$ 
are numerically checked to be positive  (see Appendix B) for values of $r$ not too
low (about $r>0.3$) and are divergent in the limit $r \rightarrow 1$.

The equation (\ref{hb2bis}) become a linear equation in $\tilde{T}(s)$ as soon
as the change of variable $z(s)=\tilde{T}(s)^{1/2}$ is performed:

\begin{equation}
2\alpha(r)s\frac{d^2}{ds^2}z(s)-2\beta(r)\frac{d}{ds}z(s)+sz(s)=0
\label{finaleq}
\end{equation}

giving the solution:

\begin{equation}
z(s)=As^{\alpha'}J_{\alpha'(r)}(\beta'(r)s)+Bs^{\alpha'(r)}N_{\alpha'(r)}(\beta'(r)s)
\label{sol}
\end{equation}

where $J_{\alpha'}$ and $N_{\alpha'}$ are the Bessel functions of the first kind and the
second kind respectively, $\alpha'(r)=(\alpha(r)+\beta(r))/(2\alpha(r))$ is real
and positive,
$\beta'(r)=(1/(2\alpha(r)))^{1/2}$ is real and is considered in its
positive determination, moreover they
present the elastic values $\alpha'(1)=1$ and $\beta'(r
\rightarrow 1)=0$ (see appendix B), while $A$ and $B$ are constants that
must be determined with assigning the boundary conditions. 

Then we can derive the expressions for
$\tilde{T}(l)$ and $\tilde{n}(l)$:

\begin{equation}
\tilde{T}(l)=(H-l)^{2\alpha'(r)}(AJ_{\alpha'(r)}(\beta'(r)(H-l))+BN_{\alpha'(r)}(\beta'(r)(H-l)))^2
\label{t_sol}
\end{equation}

\begin{equation}
\tilde{n}(l)=\frac{(H-l)^{1-2\alpha'(r)}}{(AJ_{\alpha'(r)}(\beta'(r)(H-l))+BN_{\alpha'(r)}(\beta'(r)(H-l)))^2}
\label{n_sol}
\end{equation}

To calculate the expressions of $\tilde{T}$ and $\tilde{n}$ as a function of the
original coordinate $\tilde{y}$ one needs to solve the equation 

\begin{equation}
\frac{d}{dl}\tilde{y}(l)=\frac{1}{\tilde{n}(l)}
\label{ldiy}
\end{equation}

putting in it the solution (\ref{n_sol}). However one can obtain a
comparison with the numerical simulations using the new coordinate
$l$. The main problem, at this point, is a discussion of the
boundary conditions needed to eliminate the constants $H$, $A$ and
$B$. 

One could impose that $n(l_{max})=0$ at $l_{max}=\sigma/L_x$.
From this condition immediately follows that
$H=\sigma/L_x$. A second condition can be obtained imposing a
vanishing derivative of the temperature at $l_{max}$, that is 

\begin{equation}
\left(\frac{d}{dl}\tilde{T}(l)\right)_{l=\sigma/L_x}=0
\end{equation}

The third condition is the most delicate: it must contain the rate of energy injection coming
from the vibrating wall. This rate depends
upon the parameter $T_w$ (or $A_w$ and $\omega_w$) and upon the
particles flux impinging on the wall $\Phi=n_+L_x\overline{v_+}$ where $n_+$ is the
number density of particles approaching the wall and
$\overline{v_+}$ is their velocity averaged near the wall. The
first may be simply estimated as $n_+=n(0)/2$. Moreover, if the
velocity of the macroscopic flow is zero, the average velocity of
the impinging particles is due only to fluctuations of $u$, that is 
$\overline{v_+} \approx \sqrt{k_BT(0)/m}$. In a collision with
the wall, the average energy gain is given by:

\beq \label{deltae}
\overline{\Delta E_w}=\frac{m}{2}(\overline{|{\bf v'}|^2-|{\bf
v}|^2})=\frac{m}{2}[ \overline{(v_x')^2}+\overline{(v_y')^2}-
\overline{(v_x)^2}-\overline{(v_y)^2}]
\end{equation}

which is different for the stochastic or the periodic case,
respectively:

\bea
\overline{\Delta E_{ws}}&=&\frac{3k_BT_w}{2}-k_BT(0) \\
\overline{\Delta E_{wp}}&=&\frac{m(1+r_w)^2A_w^2
\omega_w^2}{4}-\frac{(1-r_w^2)k_BT(0)}{2}
\end{eqnarray}

obtained straightforwardly from the eq. (\ref{deltae}) assuming no
correlations between the velocity of the wall and that of 
the approaching particles.

Then a non-closed expression for
the rate of energy injection coming from the wall reads:
\bea
W_{ws}&=&\overline{\Delta
E_{ws}}\Phi=\frac{3}{4}k_BT_wL_xn(0)\sqrt{\frac{k_BT(0)}{m}}-
\frac{L_xn(0)(k_BT(0))^{3/2}}{2\sqrt{m}} \\
W_{wp}&=&\overline{\Delta
E_{wp}}\Phi=\frac{m(1+r_w)^2A_w^2\omega^2}{8}L_xn(0)\sqrt{\frac{k_BT(0)}{m}}-
\frac{(1-r_w^2)}{4}L_xn(0)\frac{(k_BT(0))^{3/2}}{\sqrt{m}}
\end{eqnarray}

The above expressions are useful to establish the third needed boundary
conditions. In order to do that, they must be compared with the energy dissipation rate due 
to inelastic collisions. The local dissipation rate is given by 
$\zeta(y)k_BT(y)=C(r)\sigma n(y) (k_BT(y))^{3/2}/\sqrt{m}$ (see Appendix B). 
The instantaneous balance between energy
injection and dissipation in collisions reads, then,
 
\beq
W=\frac{1}{L_x L_y}\int_0^{L_x}dx \int_0^{L_y}dy \zeta k_BT(y)=
\frac{(-\sigma g)^{3/2}mC(r)}{L_y}\int_0^{\sigma/L_x}dl\tilde{T}(l)^{3/2}
\end{equation}

where $W$ is $W_{ws}$ or $W_{wp}$.

\begin{figure}[h]
	\centerline{\psfig{figure=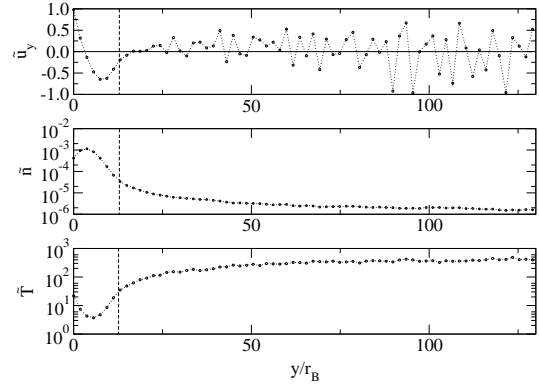,angle=-90,width=8cm}}
	\caption{Profiles of dimensionless hydrodynamic fields
	$\overline{n}$, $\overline{v}_y$ and $\overline{T}$ versus the
	dimensionless height $y/r_B$, for the model a) with the
	stochastic wall at temperature $T_w=250$. $N=5000$, $N_w \approx
	180$, $r=0.7$, $r_w=0.7$, $g_e=-1$. The dashed vertical
	line marks the same  height of figure (\ref{fig:h3}) }
	\label{fig:h2}
	\end{figure}

Apart of the difficulty of solving the boundary conditions to give an
expression of $A$ and $B$ as functions of the parameters of the model,
one must observe that the hydrodynamic description given here is
ill-posed from the beginning for what concerns a broad boundary layer near the bottom
wall. A simple look to the profiles of
$\overline{u}_y(\overline{y})=u_y(\overline{y})/\sqrt{-g_er_B}$ and
$\overline{T}(\overline{y})=T_y/(-g_er_B)$ (the overlined quantities
$\overline{n}$, $\overline{u}$, $\overline{T}$ and $\overline{y}$ are analogue of the
dimensionless variables $\tilde{n}$, $\tilde{u}$, $\tilde{T}$ and
$\tilde{y}$ with the
assumption $m=1$, $k_B=1$ and $\sigma=r_B$) in the
figure (\ref{fig:h2}) can
give the idea. We expect from the continuity equation (\ref{h1}), as
discussed above, $u_y(y)=0$ for every $y$, while a broad region
appears with a non-constant and non-monotonic behavior. Moreover, even
the profile of $T(y)$ shows an extremal point, in this case a minimum: but
from equation (\ref{heatflux}), taking in account the substitution
(\ref{substitution}), it can be seen that the imposition 
$\frac{d}{d\tilde{y}}\tilde{T}=0$ gives the following relation:

\begin{equation}
\tilde{T}(l)^{1/2}\frac{d^2}{dl^2}\tilde{T}(l)=\frac{C(r)}{B(r)-A(r)}\tilde{T}(l)^{3/2}
\label{nominimum}
\end{equation}

where the fraction $C/(B-A)=-1/\alpha$ is numerically checked to be negative
from a value of $r$ lower than $0.4$ (see appendix B). The relation (\ref{nominimum})
states that if $\tilde{T}>0$ a minimum (that is a positive value of
$\frac{d^2}{dl^2}\tilde{T}$) cannot be expected. Similar profiles for
$T(y)$, with a minimum, have been obtained in other simulations~\cite{soto}.

\begin{figure}[h]
	\centerline{\psfig{figure=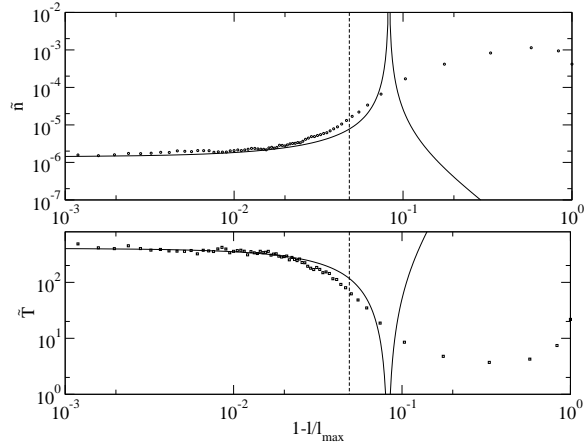,angle=-90,width=8cm}}
	\caption{Profiles of dimensionless hydrodynamic fields
	$\tilde{n}$ and $\tilde{T}$ versus $1-l/l_{max}$ (the new
	coordinate $l$ is defined in section V and
	$l_{max}=\sigma/L_x\approx r_B/L_x$), for the model a) with the
	stochastic wall at temperature $T_w=250$. $N=5000$, $N_w \approx
	180$, $r=0.7$, $r_w=0.7$, $g_e=-1$. The solid lines are
	the theoretical fit using the hydrodynamics model of Brey et
	al. The vertical dashed line marks the height (also
	appearing in figure (\ref{fig:h2})) where $\tilde{T}$ 
	presents a minimum and, therefore, goes to $0$.}
	\label{fig:h3}
	\end{figure}

A tentative fit is presented in figure
(\ref{fig:h3}). Here we used three boundary conditions obtained directly from
the simulations: a value of $nT$ at a certain height $y_1$ to obtain
directly $H$, the value of $T$ and the value of its derivative at
heights $y_2$ and $y_3$ respectively, with all $y_1$, $y_2$, $y_3$
not far from the top wall. In this tentative fit the problems
discussed above appear clearly: there is a broad region near the
bottom wall (see also figure (\ref{fig:h2})) where the theoretical
solution of the equations (\ref{hb1})-(\ref{hb2}) is qualitatively
different from the simulation data.

The qualitative inconsistencies between the observed profiles and the
hydrodynamics in a broad boundary layer near the vibrating wall are probably due to
the high density gradients present in this region. The high density
gradients represent a numerical but also a conceptual problem: it is
numerical because the profiles shown in figure (\ref{fig:h2})
are obtained by means of a coarse graining in horizontal stripes
$B(y,\Delta y)$ and so they can be compared to the theoretical profiles
only if the density in these stripes is approximately homogeneous; it
is conceptual because this hydrodynamic description is based upon the
Navier-Stokes approximation, that is an expansion of $f({\bf x}, {\bf
v},t)=f[{\bf v}|n({\bf x},t),{\bf u}({\bf x},t),T({\bf x},t)]$ up to
the first order in the gradients of the fields $n$, ${\bf u}$, $T$.

It must be stressed the fact that this boundary layer problem affects
the description of the whole system in a strong way, as its
global behavior (for example the scaling laws for the
global temperature or the center of mass height, extensively
investigated in~\cite{1d,warr,luding,kumaran}) emerges from the
balance between the bulk dissipation and the injection rate which
cannot be determined, even qualitatively, by a hydrodynamic study at
the level proposed in this paper.


\section{Conclusions}

We have studied, by means of a Direct Simulation Monte Carlo
algorithm, a model of granular flow in two different
bi-dimensional setups: the first version consists of an inclined plane
with periodic horizontal boundary condition, a top inelastic wall and
a vibrating bottom inelastic wall while gravity acts in the direction $y$
perpendicular to the vibrating wall and pointing toward it; in the second
version gravity acts in both $x$ and $y$ directions and the bottom
wall doesn't vibrate, therefore resembling a stationary flow along a
bi-dimensional channel. In both versions of the model we have found a
good agreement with the analogous experiments~\cite{kudrolli,azanza}.
In particular, the model with the vibrating wall shows strong
non-Gaussian behavior of the velocities which turns to a Gaussian
behavior if the angle of inclination is raised up (this should be an
effect of the increase of the heating rate, as the particles are more
frequently in contact with the vibrating wall). The same model
presents also evidence of different degrees of clusterization at
different heights and this is in contrast with the experimental
observation~\cite{kudrolli}. The model with gravity in both directions
and without vibrating wall shows a stationary flow in the horizontal
direction, where there are periodic boundary conditions: the profiles
of the number density $n(y)$, the $x$ component of the velocity $v_x(y)$ and
the granular temperature $T(y)$ as functions of the distance from the
bottom $y$ are in very good agreement with the experimental profiles,
showing a linear behavior in a broad region near the bottom which
corresponds to the region where the collisions dominate the
dynamics. This version of the model also shows strong evidences of
density dependent clusterization and a non-Gaussian behavior near the
bottom wall. The simplicity of the first setup has allowed us to
exactly solve the hydrodynamics equations for $n(y)$ and $T(y)$ following the
formulation of Brey et al.~\cite{brey}: however it is not possible
to obtain a matching condition between the bulk of the granular
assembly and the vibrating wall which is responsible for the injection
of energy. It seems to be an intrinsic problem of the high density
gradients observed near the bottom wall which the Navier-Stokes
approximation fails to describe. This suggests the need of a better
description of the boundary layer, which should include higher order
density and temperature gradients and also, at the level of the
kinetics, the non-Gaussian velocity statistics and the effect of
spatial correlations (clustering).


\section{Acknowledgements}

We wish to thank J. Brey, M.J. Ruiz Montero and V. Loreto for useful discussions.
This work was supported by the INFM through a PAIS grant.


\section*{Appendix A: the Bird's scheme for Monte Carlo solution of
Boltzmann Equation}

The Bird's scheme, often called Direct Simulation Monte Carlo (DSMC),
 was designed in the 1960s~\cite{bird} and its
derivation was a priori independent of the Boltzmann
equation. Recently its convergence to solutions of the Boltzmann
equation in a suitable limit has been proved~\cite{wagner},
reinterpreting it as a measure-valued stochastic process.

The Bird's scheme can be formulated as a fixed time ($\Delta t$) step ``molecular
dynamics-like'' simulation. At each time step the dynamics is
separated in two distinct processes: the independent evolution of
every particle and the collisions of near particles. The following
algorithm (for the single time step) is the one we implemented in this work, which is a
modification of the original scheme.

\begin{itemize}
\item Free flow: each particle evolves independently
following the equations of motion $\stackrel{.}{\bf x}={\bf v}$,
$\stackrel{.}{\bf v}={\bf g}$ with first order discretization.
\item Collisions: every particle $i$, 
during this time step has a probability $p_c$ of making a
collision, so that $p_c/\Delta t \propto \sigma$ (in fact the
collision cross section is proportional to the diameter of the
particles). If the particle collides, then another particle ${\bf x}_j,{\bf
v}_j$ is chosen with $|{\bf x}_i-{\bf x}_j| \leq r_B$ (we
call $r_B$ the ``Bird radius'', but it could also be thought as the ``Boltzmann
radius'') with probability  $p_{ij} \propto |{\bf v}_r|$; for the pair $i$,$j$ 
the postcollisional velocities are calculated as they were at
contact with a random choice of the collision parameter $\hat{\bf
n}$; this step is repeated for every particle. 
\end{itemize}

It is important to stress the fact that this is above all a Monte Carlo method
to solve the Boltzmann Equation (\ref{boltzeq}): in this sense
microscopic (short range) details are lost, the $N$ particles themselves
do not represent $N$  real grains of the granular assembly but carry
the space-time average information of many more particles.


\section*{Appendix B: the numerical coefficients in the hydrodynamic equations}

In section V the hydrodynamics of the first model is studied. The
equations with the transport coefficients calculated by Brey et
al.~\cite{brey} are used. The coefficients needed in our case are the
two thermal conductivities $\kappa$ and $\mu$ appearing in the
expression of the heat flux 

\begin{equation}
{\bf q}=-\kappa {\bf \nabla}(k_BT)-\mu {\bf \nabla}n
\end{equation}

and the coefficient $\zeta$ of the dissipative term

\begin{equation}
-\zeta k_BT.
\end{equation}

In reference~\cite{brey} the coefficients
are given for the case $d=3$ ($d$ is the dimension of the space). We
have taken the coefficients for $d=2$ from an unpublished (to our
knowledge) work of Brey et al.~\cite{brey2} and we have put them in
the following form:

\begin{eqnarray}
-\kappa&=&A(r)\frac{(k_BT)^{1/2}}{\sigma m^{1/2}} \\
-\mu&=&B(r)\frac{(k_BT)^{3/2}}{\sigma m^{1/2} n} \\
-\zeta&=&C(r)\frac{\sigma n(k_BT)^{1/2}}{m^{1/2}}
\end{eqnarray}

where 

\begin{eqnarray}
A(r)&=&-\kappa_1(r)\kappa_0\\
B(r)&=&-\mu_1(r)\kappa_0\\
C(r)&=&-\zeta_1(r)/\eta_0
\end{eqnarray}

and

\begin{eqnarray}
\kappa_0&=&\frac{2}{\sqrt{\pi}}\\
\eta_0&=&\frac{1}{2\sqrt{\pi}}\\
\mu_1&=&2\frac{\zeta_1(r)\left(\kappa_1(r)+\frac{c_1(r)}{4\zeta_1(r)}\right)}{\nu_1(r)-3\zeta_1(r)}\\
\kappa_1&=&\frac{1+c_1(r)}{\nu_1(r)-4\zeta_1(r)}\\
\zeta_1&=&\frac{1}{2}(1-r^2)\left(1+\frac{3}{32}c_1(r)\right)\\
\nu_1(r)&=&(1+r)\left(\frac{19}{8}-\frac{15}{8}r+\frac{1}{1024}(14-6r)c_1(r)\right)\\
c_1(r)&=&32\frac{(1-r)(1-2r^2)}{57-25r+30r^2(1-r)}.
\end{eqnarray}

\begin{figure}[h]
	\centerline{\psfig{figure=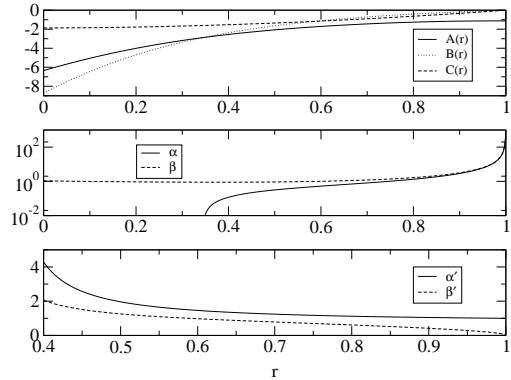,angle=-90,width=8cm}}
	\caption{Transport coefficients $A$, $B$ and dissipative
	coefficient $C$ of hydrodynamics (section V), numerical
	coefficients $\alpha$ and $\beta$ of the equation
	(\ref{finaleq}), numerical coefficients $\alpha'$ and $\beta'$
	of the solution (\ref{sol}) versus the restitution coefficient
	$r$.}
	\label{fig:coeff}
	\end{figure}

The coefficients $A(r)$, $B(r)$, $C(r)$ are plotted in the figure
(\ref{fig:coeff}). In the same figure are also presented the
coefficients $\alpha(r)=(A(r)-B(r))/C(r)$ and
$\beta(r)=(A(r)-\frac{1}{2}B(r))/C(r)$ appearing in the equation
(\ref{finaleq}) and, finally, the coefficients
$\alpha'(r)=(\alpha(r)+\beta(r))/(2\alpha(r))$ and $\beta'(r)=(1/(2\alpha(r)))^{1/2}$
appearing in the solution (\ref{sol}).


	\begin{figure}[h]
	\centerline{\psfig{figure=tflip_dens.ps,width=8cm}}
	\caption{Snapshots of the model a) with stochastic wall at
	temperature $T_w=50$ and $T_w=250$. The leftmost inset
	displays the time-averaged number density profile for both
	case. Values of other parameters: $N=500$, $N_w \approx 56$, 
        $r=0.7$, $r_w=0.7$,$g_e=-1$}
	\label{fig:tflip_dens}
	\end{figure}

\newpage

\end{document}